\documentclass[prl,twocolumn,showpacs,preprintnumbers,amsmath,amssymb,superscriptaddress,showkeys,nofootinbib]{revtex4-1}

\usepackage{epsfig}
\usepackage{amsmath}
\usepackage{graphicx}
\usepackage{mathptmx}      
\usepackage{latexsym}
\usepackage{bm}
\usepackage{comment}

\def\Xint#1{\mathchoice
{\XXint\displaystyle\textstyle{#1}}%
{\XXint\textstyle\scriptstyle{#1}}%
{\XXint\scriptstyle\scriptscriptstyle{#1}}%
{\XXint\scriptscriptstyle\scriptscriptstyle{#1}}%
\!\int}
\def\XXint#1#2#3{{\setbox0=\hbox{$#1{#2#3}{\int}$ }
\vcenter{\hbox{$#2#3$ }}\kern-.5\wd0}}

\def\dashint{\Xint{\bf-}}

\def\1{\'{\i}}
\def\Xint#1{\mathchoice
   {\XXint\displaystyle\textstyle{#1}}%
   {\XXint\textstyle\scriptstyle{#1}}%
   {\XXint\scriptstyle\scriptscriptstyle{#1}}%
   {\XXint\scriptscriptstyle\scriptscriptstyle{#1}}%
   \!\int}
\def\XXint#1#2#3{{\setbox0=\hbox{$#1{#2#3}{\int}$}
     \vcenter{\hbox{$#2#3$}}\kern-.5\wd0}}

\def\dashint{\Xint-}

\begin{document}

\title{An exact solution to the Bertsch problem \\ and the
  non-universality of the Unitary Fermi Gas}~\thanks{Supported by
  Spanish DGI (grant FIS2014-59386-P),  Junta de Andalucia (grant
  FQM225), FAPESP (grant 2016/07061-3),
CNPQ (grant 306195/2015-1) and FAEPEX (grant 3284/16).}

\author{E. Ruiz
  Arriola}\email{earriola@ugr.es}
  \affiliation{Departamento de F\'isica At\'omica, Molecular y Nuclear
  and Instituto Carlos I de Fisica Te\'orica y Computacional, Universidad de Granada, E-18071 Granada, Spain}

\author{S. Szpigel}\email{szpigel@mackenzie.br}
\affiliation{
Centro de R\'adio-Astronomia e Astrof\1sica Mackenzie, Escola de Engenharia, Universidade Presbiteriana Mackenzie, Brazil}

\author{V. S. Tim\'oteo}\email{varese@ft.unicamp.br}

\affiliation{Grupo de \'Optica e Modelagem Num\'erica (GOMNI), Faculdade de Tecnologia, Universidade Estadual de Campinas - UNICAMP, Brazil}

\date{\today}

\begin{abstract}
We analyze the universality of the Unitary Fermi Gas in its ground
state from a Wilsonian renormalization point of view and compute the
effective range dependence of the Bertsch parameter $\xi$ exactly. To
this end we construct an effective block-diagonal two-body separable
interaction with the Fermi momentum as a cut-off which reduces the
calculation to the mean field level. The interaction is separable in
momentum space and is determined by Tabakin's inverse scattering
formula.  For a vanishing effective range we get $\xi = \frac{176}{9
  \pi }-\frac{17}{3} = 0.56$. By using phase-equivalent similarity
transformations we can show that there is a class of exact solutions
with any value in the range $ 0.56 \ge \xi \ge -1/3$.
\end{abstract}

\maketitle

During the 10th Conference on Advances in Many- Body Theory, that took
place in Seattle in 1999, G. F. Bertsch posed the following challenge
(see \cite{Baker:1999np}):

{\it What are the ground state properties of the many-body system
  composed of spin-1/2 fermions interacting via a zero range, infinite
  scattering-length contact interaction?.  It may be assumed that the
  interaction has no two-body bound states. Also, the zero range is
  approached with finite ranged forces and finite particle number by
  first taking the range to zero and then the particle number to
  infinity.}

The Unitary Fermi Gas was proposed as a simple, universal and scale
invariant dilute system where interactions at low energies are strong
ranging from atoms in the ultracold regime to nuclear systems such as
neutron stars~\cite{Drut:2012md,Carlson:2012mh}. More specifically, if
$\alpha_0$ is the scattering length, $r_0$ the effective range and
$k_F$ the Fermi momentum, the unitary limit is meant to describe
systems in the range $ 1/\alpha_0 \ll k_F \ll 1/r_0 $. The interaction
is characterized by an isotropic scattering amplitude given by 
\begin{eqnarray}
f= \sum_{l=0}^\infty (2 l+1) \frac{e^{i\delta_l} \sin \delta_l}{k}
P_l(\cos \theta)\equiv \frac1{-\frac1{\alpha_0}+\frac12 r_0 k^2 -i k}\, ,
\label{eq:scat}
\end{eqnarray}
with $P_l(\cos \theta)$ Legendre polynomials and $\delta_l(k)$ the
phase shifts,
\begin{eqnarray}
\delta_0 (k)&=& \cot^{-1} \left(-\frac{1}{\alpha_0 k}+ \frac12 r_0 k \right)  ~, \nonumber \\ 
\delta_l (k) &=& 0 \qquad  {\rm for} \qquad l \ge 1 ~ . 
\label{eq:bertsch}
\end{eqnarray}
For $ \alpha_0 \to -\infty$ and $ r_0 \to 0$ one has $\delta_0(k)=
\pi/2$. In this limit, for a Fermi system with two (spin) species the
Fermi momentum $k_F$ is the only dimensionful quantity and hence the
total energy per particle should be proportional to the energy of the
free Fermi Gas:
\begin{eqnarray}
\frac{E}{N} = \xi \frac{3 k_F^2}{10 M} ~,
\end{eqnarray}
where $\xi$ is the Bertsch parameter which is expected to be a {\it
  universal number}, and in the $r_0 \neq 0$ case a {\it universal
  function} of the combination $r_0 k_F$.

This apparently simple problem provides an example of a strongly
correlated fermion system and has been a major theoretical and
experimental challenge over the last two decades.  Experimental
measurements on ultracold atomic gases for a vanishing effective
range, $r_0=0$, yield~\cite{navon2010equation} ($\xi=0.41(1)$),
\cite{luo2009thermodynamic} ($\xi=0.39(2)$), \cite{ku2012revealing}
($\xi=0.376(5)$) and \cite{zurn2013precise} ($\xi=0.37(1)$). In
Ref.~\cite{Drut:2012md} over 50 different calculations based on
different many body simulations and experiments are listed and, with a
few exceptions $\xi \sim 0.37-0.40$. The numerical resemblance
suggests, as it was tacitly expected in those studies, that this is a
universal quantity which is uniquely determined by the conditions
spelled out originally by Bertsch's challenge. Here, we provide an
exact solution of the problem and show, contrary to this general and
widespread belief, that more information is needed than assumed
hitherto.

From a theoretical point of view the strategy to deal with the Unitary
Fermi Gas has been a two-step approach: one first tunes a two-body
interaction to fulfill the Bertsch scattering condition,
Eq.~(\ref{eq:scat}), and then uses it to solve the many-body
problem. These two stages are usually discussed separately in the
literature. A prototype calculation is the one pursued recently by
Conduit and Schonberg~\cite{schonenberg2017effective} where, for any
$k_F$, they have tuned a potential to Eqs.~(\ref{eq:bertsch}) and
considered a Monte Carlo simulation up to $N=294$ particles and verify
the expected ${\cal O} (N^{-1})$ trend to the extrapolated $N=\infty$
thermodynamic limit. They find $\xi=0.388(1)$, the most precise
determination to date, and consider also finite effective range values
in the interval, $r_0 k_F \in [-2,2]$.

In the present letter we advocate for a different strategy.  Indeed,
there is a well-known and inherent ambiguity associated to this
approach; one can undertake a phase-equivalent unitary transformation
of the potential~\cite{Ekstein:1960xkd} (see e.g.
\cite{srivastava1975off} for a review). Here we propose to take
advantage of this arbitrariness by building an interaction in
block-diagonal form where the separation scale in the momentum,
$\Lambda$, is actually chosen to coincide with the Fermi momentum,
$\Lambda=k_F$. This has the important consequence that the mean field
result already yields the {\it exact} many-body solution and a
suitable choice permits to construct an analytic solution without
violating any of the conditions originally spelled out by Bertsch.

For our purposes it is convenient to formulate the two-body scattering
problem for two identical particles of mass $M$ in momentum space for
the kinematics
$$(\vec P + \vec p/2,\vec P-\vec p/2) \to (\vec P + \vec p'/2,\vec P-\vec p'/2) $$ where $\vec P$ is the
(conserved) CM momentum and $\vec p$ and $\vec p'$ the relative
momenta before and after the collision respectively. The
Lippmann-Schwinger equation, which in operator form reads $T(E) = V +
V G_0 (E) T (E)$ with $G_0(E)=(E-H_0)^{-1}$, becomes~\cite{landau2008quantum},
\begin{eqnarray}
\langle \vec p' | T (E) | \vec p \rangle = 
\langle \vec p' | V | \vec p \rangle + \int \frac{d^3 q}{(2\pi)^3}
\frac{\langle \vec p' | V | \vec q \rangle \langle \vec q | T (E) | \vec p \rangle  }{E- q^2 / 2 \mu + i \epsilon} \, , 
\label{eq:LS} 
\end{eqnarray}
where $\mu=M/2$ is the reduced mass.  In the partial wave basis,
\begin{eqnarray}
\langle \vec p' | T (E) | \vec p \rangle = \frac{8 \pi^2}{\mu}
\sum_{l m_l} Y_{l m_l} (\hat p) Y_{l m_l} (\hat p')^* T_l (p',p,E)   \, , 
\end{eqnarray}
with $Y_{l m_l} (\hat p)$ spherical harmonics and similarly for the
potential $V_l$. In terms of the means of the  $K-$matrix
fulfilling, $T_l = K_l /(1+i \sqrt{2\mu E} K_l)$ which half-off-shell
fulfills 
\begin{eqnarray}
K_l (p',p)  = V_l (p',p) + \frac{2}{\pi} \dashint_0^\infty dq \frac{q^2 V_l (p',q)}{p^2-q^2} 
K_l (q,p) \, .
\label{eq:LS-K} 
\end{eqnarray}
where $\dashint$ stands for the principal value integral and $K_l
(p',p) \equiv K_l(p',p,E=p^2/2\mu)$.  The relation with the
phase-shifts in Eq.~(\ref{eq:scat}) follows from $- 4 \pi f = \langle
\vec p' |T(E) | \vec p \rangle $ and is given by
\begin{eqnarray}
\frac{\tan \delta_l(p)}{p}=-K_l (p,p)  ~ .
\label{eq:K-mom-bare} 
\end{eqnarray}
Clearly, the conditions in Eq.~(\ref{eq:bertsch}), are fulfilled by
taking
\begin{eqnarray}
V_l(p,p')=0 \, , \quad {\rm for} \quad  l \ge 1 \, . 
\end{eqnarray}
The problem is then to find the s-wave interaction $V_0(p'p)$ from
Eq.~(\ref{eq:bertsch}), as we will discuss shortly, after reviewing
our many-body setup.

Following the conventional strategy, once our effective interaction
$V_0(p',p)$ has been tuned to the Bertsch renormalization condition,
Eq.~(\ref{eq:bertsch}), we turn now to the many body problem.  We will
work first at lowest order in perturbation theory which corresponds to
the mean field (Hartree-Fock) level, since this already provides an
upper variational estimate for {\it any} $V_0(p',p)$.  
For a two-fermion species the energy per particle at the
Hartree-Fock level, is given by~\cite{fetter1971quantum}
\begin{eqnarray}
\frac{E}{N} &=& \frac{3 k_F^2}{10 M} + 
\frac2\pi \frac4{M}\int_0^{k_F} k^2 dk \left(1-\frac{3k}{2k_F}+\frac{k^3}{2 k_F^3} \right) V_0(k,k) \nonumber \\ 
&+& {\cal O} (V^2) ~ .
\label{eq:mf}
\end{eqnarray}
According to the standard variational argument, first order
perturbation theory provides an upper bound for the true ground
state.  If we have $H=H_0+V$ and $H_0 \psi_n^{(0)} = E_n^{(0)}
\psi_n^{(0)} $, then for {\it any} normalized state $\varphi$ we have
$E_0 \le \langle \varphi | H_0 + V | \varphi \rangle $ so that taking
$\varphi= \psi_n^{(0)}$ a Slater determinant leading to
Eq.~(\ref{eq:mf}). $ E_0 \le E_n^{(0)} + \langle \psi_n^{(0)} | V |
\psi_n^{(0)} \rangle $. The neglected higher order corrections
correspond to transitions $\vec p \to \vec p'$ above the Fermi-level,
$|\vec P \pm \vec p/2| \le k_F \le |\vec P \pm  \vec p'/2| $ which requires
$p'> k_F > p$.  Note, however, that if we have $V_0 (p',p)=0$ for $p'
> k_F > p$ higher order corrections {\it vanish identically}.

The main ingredient in our construction is thus to separate the
two-body (relative) Hilbert space into two orthogonal (and decoupled)
subspaces ${\cal H}= {\cal H}_P \oplus {\cal H}_Q$ which are below or
above some given $\Lambda$ respectively. This can be denoted by
projection operators $P$ and $Q$, fulfilling $P^2=P $ and $Q^2=Q $ and
$P Q= QP=0 $ and $P+Q=1$.  This separation endows the Hamiltonian $H$
with a block structure, which can equivalently be transformed by a
unitary transformation $U$ into a block-diagonal form
\begin{eqnarray}
H = \begin{pmatrix}
 P H P       &  P H Q \\ \\
    Q H P   &  Q H Q 
  \end{pmatrix} = U \begin{pmatrix}
 P H_\Lambda P       &  0 \\ \\
    0   &  Q \bar H_\Lambda Q 
  \end{pmatrix} U^\dagger 
\end{eqnarray}
where $H_\Lambda$ and $\bar H_\Lambda$ describe the low energy and
high energy dynamics respectively. Of course, we can split the
Hamiltonian as $H=T+V$, with $T$ and $V$ kinetic and potential
energies respectively. For the case when $U$ commutes with the kinetic
energy, $[U,T]=0$, an equivalent decomposition holds for the potential
$V$ in terms of $V_\Lambda$ in the P-space and $\bar V_\Lambda$ in the
Q-space.~\footnote{There are many inequivalent ways how this procedure
  can be carried out as a result of a finite number of steps. A
  particular implementation is to achieve Block-Diagonalization in a
  continuous way  in terms of flow
  equations~\cite{Anderson:2008mu}. We will exploit this freedom
  below.}  Our idea is to assume already the former decomposition to
the two-boby problem from the start and to consider the following
potential in the momentum P-space
\begin{eqnarray}
V_\Lambda (p',p) &=& \theta(\Lambda-p') \theta(\Lambda-p) v(p',p) \, .
\label{eq:pseudo}
\end{eqnarray}
The effective interation $V_{\Lambda}(p',p)$ depends explicily on the
separation scale or cut-off $\Lambda$. It corresponds to a
self-adjoint operator, $V_\Lambda(p',p) = V_\Lambda(p,p')^*$, acting
in a reduced model Hilbert space with $p,p' \le \Lambda$.
Due to the fact that the transformation is unitary we get  that the phase-shift 
associated with $V_{\Lambda}(p',p)$ is just 
\begin{eqnarray}
\delta_{0,\Lambda} (p) = \delta_0 (p) \Theta(\Lambda-p) \, , 
\label{eq:del-lowk}
\end{eqnarray}
in the $P$-model space. Note that for $\Lambda=k_F$, the $Q$-space
becomes {\it irrelevant} in the many body problem, so that
\begin{eqnarray}
\xi &=& 1 + \frac{80}{3 \pi k_F^2}\int_0^{k_F} k^2 dk
\left(1-\frac{3k}{2k_F}+\frac{k^3}{2 k_F^3} \right) V_{k_F}(k,k) \, .
\label{eq:mf-exact}
\end{eqnarray}
We stress that for this choice of two body potential with only the
s-wave contribution this equation is {\it exact}. We are only left
with the determination of a suitable function $v(p',p)$.

\begin{figure}[tbh]
\begin{center}
\includegraphics[height=8cm,width=8cm]{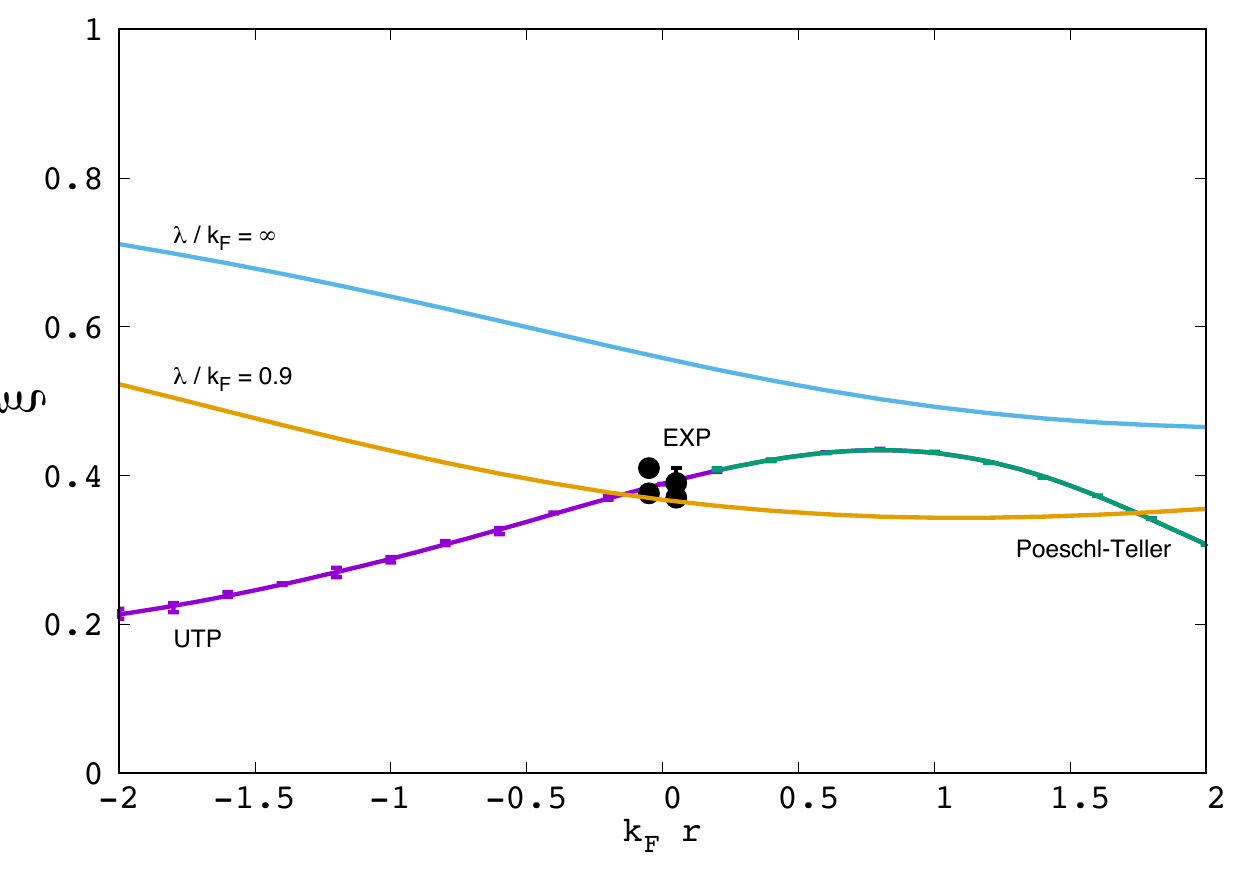}
\end{center}
\caption{Exact Bertsch parameter $\xi$ as a function of the effective
  range $r_0$ in units of the Fermi momentum $k_F$ for $\alpha_0 \to
  -\infty$ for a separable potential from the Tabakin's solution (
  $\lambda/k_F=\infty$) and its SRG evolution to $\lambda/k_F=0.9 $
  (see main text). We also compare to the Monte Carlo simulation of
  Ref.~\cite{schonenberg2017effective} using UTP and Poeshl-Teller
  interaction, and the experimental measurements on atomic gases for
  $r=0$~\cite{navon2010equation} ($\xi=0.41(1)$),
  \cite{luo2009thermodynamic} ($\xi=0.39(2)$), \cite{ku2012revealing}
  ($\xi=0.376(5)$) and \cite{zurn2013precise} ($\xi=0.37(1)$).}
\label{fig:pxp'}
\end{figure}

Within the class of solutions given by Eq.~(\ref{eq:pseudo}) we can
still exploit the arbitrariness to choose an interaction which will
provide an {\it analytical} solution to the Bertsch's problem.  Here,
we will search for a {\it separable} interaction solution of the form
\begin{eqnarray}
V_0 (p',p) = \pm g(p') g(p) \, . 
\label{eq:sep}
\end{eqnarray}
The $\pm$ sign specifies a repulsive and an attractive interaction
respectively.  This approach will work up to a value of $p < \Lambda$
where the phase shift $\delta_\Lambda (p)$  does reproduce
Eq.~(\ref{eq:bertsch}).  For a separable potential of the form of
Eq.~(\ref{eq:sep}) the solution of the Lippmann-Schwinger equation
reads reads~\cite{Arriola:2014nia,Arriola:2016fkr}
\begin{eqnarray}
 p \cot \delta_0 (p) &=& - \frac{1}{V_0 (p,p)} \left[1- \frac{2}{\pi}
   \dashint_0^\infty dq \frac{q^2}{p^2-q^2} V_0 (q,q) \right] \nonumber
 \\ &=& -\frac1{\alpha_0} + \frac12 r_0 p^2 
\label{eq:ERE}
\end{eqnarray}
For separable potentials the inverse scattering problem may be solved
in quadrature by the Tabakin's formula devised in
1969~\cite{Tabakin:1969mr} (for a review see
e.g. \cite{chadan2012inverse}). In our case the attractive solution
{\it without} bound state will be the pertinent one, reading
\begin{eqnarray}
[g(k)]^2= \frac{\sin \delta (k)}{k}\exp \left[ -\dashint_{-\infty}^\infty \frac{\delta(k')}{k-k'} dk'\right] \, , 
\label{eq:tabakin}
\end{eqnarray}
where $\delta(-k) = - \delta(k)$ and the real and positive $g(k)$ is
taken provided $\sin \delta(k)>0$ or $0 \le \delta(k) \le \pi$, a
condition fulfilled by Eq.~(\ref{eq:bertsch}) for {\it any} value of
$\alpha_0$ and $r_0$. In our case. according to
Eq.~(\ref{eq:del-lowk}) we have a limited integration interval
$-\Lambda \le k \le \Lambda$. The case with $\alpha_0 \to - \infty$
and $r_0=0$ can be worked out explicitly, yielding for $p>0$
\begin{eqnarray}
g(p) = \frac{\theta(\Lambda-p)}{\sqrt[4]{\Lambda^2 - p^2 }} \, .
\end{eqnarray}
For $\Lambda=k_F$, the potential satisfying {\it exactly} the
conditions is
\begin{eqnarray}
V_{k_F}(k',k) = - \frac{\theta(k_F-k')}{\sqrt[4]{k_F^2 - k'^2 }}
\frac{\theta(k_F-k)}{\sqrt[4]{k_F^2 - k^2 }} \, , 
\label{eq:pot-sol}
\end{eqnarray}
which as expected depends on $k_F$. It can be readily checked that for
$k>0$ one has
\begin{eqnarray}
\delta_0(k) = \frac{\pi}2 \theta(k_F-k) \, . 
\end{eqnarray}
A direct
evaluation of the integral in Eq.~(\ref{eq:mf}) yields
\begin{eqnarray}
\xi=\frac{176}{9 \pi }-\frac{17}{3} = 0.558 \, , 
\end{eqnarray}
contrary to the ``universal'' value
$\xi=0.37-0.40$~\cite{Drut:2012md}. The case $\alpha_0 \to -\infty $
and $r_0 \neq 0$ can also be computed analytically from Tabakin's
formula, Eq.~(\ref{eq:tabakin}), in terms of dilogarithmic functions.
The result is depicted in Fig.~\ref{fig:pxp'}, contradicting Conduit
and Schonberg~\cite{schonenberg2017effective}.

We note, however, that the previous result, while exact is not unique.
As already mentioned, one can still undertake a further
phase-equivalent unitary transformation within the $P-$space of the
potential $V \to U^\dagger V U$ which preserves the essential feature
that the interaction does not allow transitions above the Fermi
surface, but reshuffles the $v(p',p)$ function.

\begin{figure}[tbh]
\begin{center}
\includegraphics[height=8cm,width=8cm]{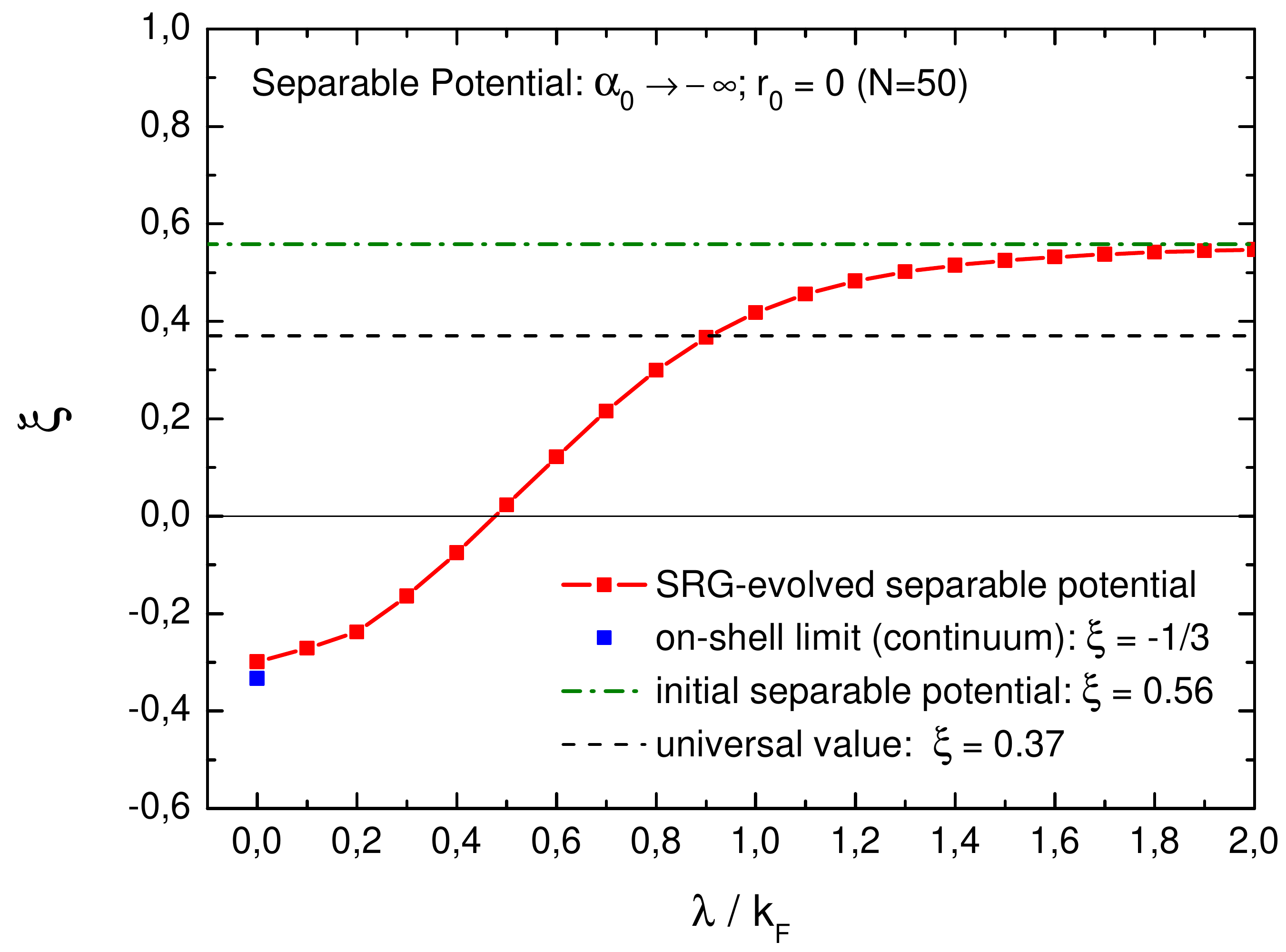}
\end{center}
\caption{Unitary Phase-equivalent evolution of the Bertsch
  parameter $\xi$ as a function of the similarity cut-off $\lambda
  \equiv 1/s^{1/4}$ (squares) for a discrete momentum grid with $N=50$
  points. For $N=\infty$ we mark the
  initial separable potential corresponding to $\xi=0.56$ and the
  $\lambda \to 0$ final potential corresponding to the
  $\xi=-1/3=-0.33$.  The horizontal line corresponds to the
  ``universal'' value $\xi=0.37$ obtained by many calculations and
  experiments~\cite{Drut:2012md}.}
\label{fig:xi-simcut}
\end{figure}

A simple way to generate a continuous one-parameter unitary
transformation according to the previous requirements is by means of
the Similarity Renormalization Group (SRG) method introduced by Wilson
and Glazek~\cite{Glazek:1994qc} (for a review see e.g.
\cite{kehrein2006modern}). Defining the Hamiltonian $H_s=T+V_s$ at the
operator level the SRG equation with the Wilson, $G_s=T$, generator reads
\begin{eqnarray}
\frac{d V_s}{ds} = \left[ [T, V_s ], T+V_s \right] \, . 
\label{eq:WG}
\end{eqnarray}
This evolution equation monotonously minimizes the Frobenius norm of
the potential $|| V_s ||^2 = {\rm tr} ( V_s^2 )$, since $ d {\rm tr} (
V_s^2 ) / ds < 0$ and for $s \to \infty $ provides $\lim_{s\to
  \infty}[T,V_s]=0$, i.e. the potential energy operator becomes
diagonal in momentum space and hence
on-shell~\cite{Arriola:2013gya,Timoteo:2016vlp} . Here, one has for
$v(p,p')$ in Eq.~(\ref{eq:pseudo}) and in our case ($\Lambda=k_F$) the
following equation,
\begin{eqnarray}
\frac{d~v_s(p,p')}{ds} &=& - (p^2-p'^2)^2~v_s(p,p') +\frac{2}{\pi} \int_0^{k_F} dq~q^2\nonumber \\
 &\times& (p^2 + p'^2 - 2 q^2)~v_s(p,q)~v_s(q,p') \, , 
\label{eq:srg}
\end{eqnarray}
where $s=1/\lambda^4$ and $\lambda$ is the similarity cutoff. The flow
equation generates a set of isospectral interactions that approaches a
diagonal form as $s\to\infty$ (or $\lambda\to0$). Only in few cases
have the SRG integro-diferential equations been solved
analytically~\cite{Szpigel:1999gf}.  Their numerical treatment
requires introducing a finite momentum grid, so that results in the
continuum are taken as a limiting
procedure~\cite{Arriola:2014nia,Arriola:2016fkr}. Taking the
$v_s(k,k)$ into the mean field energy one obtains a phase-equivalent
flow equation for the Bertsch parameter, $\xi_s$. Defining $\varphi(x)=
1-3x/2+x^3/2 $, Eq.~(\ref{eq:srg}) yields the inequality
\begin{eqnarray}
  \frac{d \xi_s }{ds} &=&  \frac{80}{3 \pi} \left (\frac{2}{\pi} \right)^2 \int_0^{k_F} dq~q^2 \int_0^{k_F}dk~k^2  \left[\left(\frac{k}{k_F} \right)^2-\left(\frac{q}{k_F} \right)^2
\right]
 \nonumber \\
& \times &  
\left[\varphi\left(\frac{k}{k_F} \right)- \varphi\left(\frac{q}{k_F} \right)
\right]~|v_s(k,q)|^2  \le 0 \, , 
\label{eq:xip}
\end{eqnarray}
since $\varphi(x)$ is a decreasing function,
$\varphi'(x)=-3(1-x^2)/2<0$, and thus $(x^2-y^2)
[\varphi(x)-\varphi(y)] < 0$ in $0 < x,y < 1$. This inequality
actually shows that the Bertsch parameter is not determined uniquely
from the s-wave phase-shift and hence $\xi$ {\it is not universal}.

In the on-shell limit, $s\to\infty$ ($\lambda \to 0$), $v_s(p',p)$ becomes
diagonal, and one has thus $d \xi_s /ds \to 0$. The limiting value of
Eq.~(\ref{eq:WG}) was determined in terms of the scattering
phase-shifts~\cite{Arriola:2014nia,Arriola:2016fkr}. Adapted to our
Eq.~(\ref{eq:srg}) and in the absence of bound states, the limit
becomes a fixed point. If $k' \neq k$ then
\begin{eqnarray}
  \lim_{s \to \infty } v_s (k,k) = -\frac{\delta_0(k)}k \, , \quad 
  \lim_{s \to \infty } v_s (k',k) = 0 \quad 
 \, , 
  \label{eq:asympv}
\end{eqnarray}
which is asymptotically stable~\cite{Arriola:2014nia,Arriola:2016fkr}
and the solutions are attracted to this one.  Hence, for $\delta_0
(k)=\pi/2$ and computing a trivial integral we finally get a fixed
point solution
\begin{eqnarray}
\lim_{s \to \infty} \xi_s =  1 - \frac{4}{3}=  - \frac13 \, . 
\end{eqnarray}
This corresponds to a unstable system. Thus, the previous argument
shows that regardless of the initial function $v(p',p)$ at $s=0$ with
a given value of $\xi$, there is a phase-equivalent potential where
$\xi < 0$.  

The fixed point solution only depends on the choice $\Lambda=k_F$. In
the case, $k_F < \Lambda$, the mean field result is only an upper
bound, $\xi \le \xi_{\rm MF}$. The flow equations for $v_s(p',p)$ and
$\xi_{\rm MF}$ read as Eq.~(\ref{eq:srg}) and Eq.~(\ref{eq:xip})
respectively with the replacement $\int_0^{k_F} dq \to
\int_0^{\Lambda} dq $, so that the same inequality holds. Thus, one
has $\xi_s \le \xi_{\rm MF} \to - 1/3 $. In the general case, with
finite $\alpha_0 \neq 0$ and $r_0 \neq 0$, see Eq.~(\ref{eq:bertsch}),
the sign of $\lim_{s \to \infty} \xi_s$ depends on their particular
values. For instance for $r_0=0$, the intercept $\xi_\infty = 0$
happens for $\alpha_0 k_F =-7.5378$ and for $\alpha_0 \to -\infty$ one
has $ r_0 k_F= 2.038$.

As we have mentioned above, these solutions, while exact, are not
unique; we can still carry out a phase-equivalent transformation and
change the value of $\xi$. In particular, from the SRG equations on a
momentum grid~\cite{Arriola:2014nia,Arriola:2016fkr} we can cover
continuously all values from the starting one to the final
one~\footnote{ We take $N=50$ points and Gauss-Legendre points. From
  the discretized form of Eq.~(\ref{eq:mf-exact}) we get $\xi=
  0.55806$ using Eq.~(\ref{eq:pot-sol}) and $\xi=-0.298993$ for
  Eq.~(\ref{eq:asympv}). The SRG equations become increasingly stiff
  for large $N$'s and small
  $\lambda$'s~\cite{Arriola:2014nia,Arriola:2016fkr,Timoteo:2016vlp}. We
  estimate an error in $\xi$ about 0.03, which is compatible with the
  expected flat behaviour at the fixed point.}. This is shown in
Fig.~\ref{fig:xi-simcut} as a function of $\lambda/k_F$.

In particular, we could tune the SRG-scale $\lambda$ to obtain from
the potential $V_{k_F}(k',k)$ given by Eq.~(\ref{eq:pot-sol}) with
$\xi=0.558$ the ``universal'' value $\xi=0.37-40$ obtained in many
calculations and experiments~\cite{Drut:2012md}. We find that
$\xi=0.37$ happens for $\lambda /k_F = 0.9$, see
Fig.~\ref{fig:xi-simcut}. It is of course tempting to analyze the
effective range behaviour at the scale $\lambda/k_F=0.9$. This is done
in Fig.~\ref{fig:pxp'} and compared again with the recent Monte Carlo
calculation of Conduit and
Schonberg~\cite{schonenberg2017effective}. As we see the lack of
universality of $\xi$ is reinforced for finite $r_0$ even after tuning
the $r_0=0$ value.

Besides illustrating the lack of universality our findings provide
quite different values showing that the numerical resemblance among
the many calculations and experiments is due to a common, yet unknown,
feature among them which was not spelled out in the famous Bertsch's
problem and deserves an explanation. We can think of several reasons
for not reproducing neither the Monte Carlo calculation nor the
experimental data on ultracold atoms which agree among
themselves. Firstly, Monte Carlo calculations have only been carried
out for {\it local} potentials. The solution of the inverse scattering
problem exists~\cite{chadan2012inverse} and will be discussed
elsewhere. Secondly, the potentials experienced between neutral atoms
are van der Waals-like and hence local. Thus, we conjecture that
locality is the additional condition underlying the observed
universality. Work along these lines is in progress.


\end{document}